\documentclass[12pt]{article}

\pdfoutput=1

\usepackage{cite}

\usepackage{amssymb,latexsym}

\usepackage{epstopdf}

\usepackage{graphics,epsfig}

\usepackage{amsmath,amsfonts}
\usepackage{amssymb}
\usepackage{float}
\usepackage{amssymb,latexsym}

\usepackage{appendix}

\usepackage{graphicx}
\DeclareGraphicsExtensions{.pdf,.png,.jpg}

\usepackage{color}

\usepackage{mathrsfs}

\DeclareMathAlphabet{\mathpzc}{OT1}{pzc}{m}{it}

\let\a=\alpha \let\b=\beta \let\g=\gamma \let\d=\delta \let\e=\epsilon
\let\z=\zeta  \let\th=\theta  \let\k=\kappa
\let\l=\lambda \let\m=\mu \let\n=\nu \let\x=\xi \let\p=\pi %\let\r=\rho
\let\s=\sigma   \let\f=\phi  
 
\let\w=\omega      \let\G=\Gamma \let\D=\Delta \let\Th=\Theta \let\L=\Lambda
\let\X=\Xi  \let\S=\Sigma  \let\Y=\Psi
 
\let\la=\label  
  
\def\nn{\nonumber} \def\bd{\begin{document}} \def\ed{\end{document}}
\def\ds{\documentstyle} \let\fr=\frac \let\bl=\bigl \let\br=\bigr
\let\Br=\Bigr \let\Bl=\Bigl
\let\bm=\bibitem
\let\na=\nabla
\def\tU{{\widetilde U}}
\let\pa=\partial \let\ov=\overline
\def\ie{{\it i.e.\ }}
\newcommand{\be}{\begin{equation}}
\newcommand{\ee}{\end{equation}}
\def\ba{\begin{array}}
\def\ea{\end{array}}
\def\ft#1#2{{\textstyle{{\scriptstyle #1}\over {\scriptstyle #2}}}}
\def\fft#1#2{{#1 \over #2}}
\def\F#1#2{{ F_{#1}^{(#2)} }}
\def\cF#1#2{{ {\cal F}_{#1}^{(#2)} }}

\def\R{{\bf R}}
\def\sst#1{{\scriptscriptstyle #1}}
\def\oneone{\rlap 1\mkern4mu{\rm l}}
\def\e7{E_{7(+7)}}
\def\td{\tilde}
\def\wtd{\widetilde}
\def\im{{\rm i}}
\def\bog{Bogomol'nyi\ }
\newcommand{\ho}[1]{$\, ^{#1}$}
\newcommand{\hoch}[1]{$\, ^{#1}$}
\newcommand{\bea}{\begin{eqnarray}}
\newcommand{\eea}{\end{eqnarray}}
\newcommand{\ra}{\rightarrow}
\newcommand{\lra}{\longrightarrow}
\newcommand{\Lra}{\Leftrightarrow}
\newcommand{\ap}{\alpha^\prime}
\newcommand{\bp}{\tilde \beta^\prime}
\newcommand{\cB}{{\cal B}}
\newcommand{\cO}{{\cal O}}
\newcommand{\vecx}{\vec{x}}
\newcommand{\vecy}{\vec{y}}
\newcommand{\vecp}{\vec{p}}
\newcommand{\vecq}{\vec{q}}
\newcommand{\tr}{{\rm tr} }
\newcommand{\Tr}{{\rm Tr} }
\newcommand{\NP}{Nucl. Phys. }

\newcommand{\cL}{{\cal L}}
\newcommand{\cA}{{\cal A}}
\newcommand{\cT}{{\cal T}}
\newcommand{\cD}{{\cal D}}
\newcommand{\cH}{{\cal H}}
\newcommand{\cR}{{\cal R}}

\def\sst#1{{\scriptscriptstyle #1}}
\def\0{{\sst{(0)}}}
\def\1{{\sst{(1)}}}
\def\2{{\sst{(2)}}}
\def\3{{\sst{(3)}}}
\def\4{{\sst{(4)}}}
\def\5{{\sst{(5)}}}
\def\6{{\sst{(6)}}}
\def\7{{\sst{(7)}}}
\def\8{{\sst{(8)}}}
\def\9{{\sst{(9)}}}
\def\p{{\sst{(p)}}}
\def\q{{\sst{(q)}}}
\def\ve{\varepsilon}
\def\vf{\varphi}
\def\F{\Phi}
\def\wg{\wedge}

\def\thb{\bar{\theta}}
\def\Thb{\bar{\Theta}}
\def\barp{\bar{p}}
\def\barq{\bar{q}}
\def\barc{\bar{c}}
\def\bard{\bar{d}}
\def\e{\epsilon}

\def \bi{\bibitem}
\def \la {\label}

\def \l {\lambda}
\def\foot{\footnote}
\def \tl  {{\tilde \l}}
\def \sql {{\sqrt \l}}
\def \adss {$AdS_5 \times S^5$\ }
\newcommand{\rf}[1]{(\ref{#1})}
\def \ov {\over}

\def\th{\theta}
\def\Th{\Theta}
\def\vth{\vartheta}
\def\btheta{{\bar\theta}}
\def\ttheta{{{\tilde\theta}}}
\def\bttheta{{{\bar\ttheta}}}
\def\vth{\vartheta}

\def\ra{\rightarrow}
\def\N{\nabla}
\def\F{{\cal F}}
\def\uM{\underline{M}}
\def\uA{\underline{A}}
\def\uN{\underline{N}}
\def\uP{\underline{P}}
\def\ua{\underline{a}}
\def\ub{\underline{b}}
\def\uc{\underline{c}}
\def\ud{\underline{d}}
\def\ue{\underline{e}}
\def\uf{\underline{f}}
\def\ui{\underline{i}}
\def\uj{\underline{j}}
\def\uk{\underline{k}}
\def\ul{\underline{l}}
\def\ual{\underline{\alpha}}
\def\ube{\underline{\beta}}
\def\um{\underline{m}}
\def\un{\underline{n}}
\def\up{\underline{p}}
\def\uq{\underline{q}}
\def\ur{\underline{r}}
\def\us{\underline{s}}
\def\umu{\underline{\mu}}
\def\unu{\underline{\nu}}
\def\ula{\underline{\l}}
\def\uka{\underline{\k}}
\def\usi{\underline{\s}}
\def\urh{\underline{\r}}
\def\cc{\circ}
\def\eqv{\equiv}

\def\ni{\noindent}

\def\Ep{E^{{}^{(+)}}}
\def\Em{E^{{}^{(-)}}}

\def\Mp{M^{{}^{(+)}}}
\def\Mm{M^{{}^{(-)}}}

\def \ha{{1\ov 2}}

\def\r{\rho}

\def\Y{{\rm Y}}
\def\X{{\rm X}}
\def\tY{\tilde{\rm Y}}
\def\tX{\tilde{\rm X}}
\def\dY{\dot{\rm Y}}
\def\dX{\dot{\rm X}}

\def \J {\mathcal{J}}
\def \del {\partial}

\def\dF{\dot{F}}
\def\dG{\dot{G}}
\def\df{\dot{f}}
\def \E {{\cal E}}
\def \S {{\cal S}}
\def \J {{\cal J}}

\def\ms{\mathcal{S}}
\def\mj{\mathcal{J}}
\def\soj{\fr{\ms}{\mj}}
\def \R {{\bf R}}
\def \om {\omega}
\def \bE {\bar E}
\def \x {{\cal X}}

\def \bi{\bibitem}
\def \la {\label}

\def \l {\lambda}
\def\foot{\footnote}
\def \tl  {{\tilde \l}}
\def \sql {{\sqrt \l}}
\def \adss {$AdS_5 \times S^5$\ }
\def \ov {\over}

\def \varpi {{\rm w}}

\def\thb{\bar{\theta}}
\def\Thb{\bar{\Theta}}
\def\mb{\bar{\m}}
\def\ab{\bar{\a}}
\def\zb{\bar{z}}
\def\psib{\bar{\psi}}
\def\barp{\bar{p}}
\def\barq{\bar{q}}
\def\barc{\bar{c}}
\def\bard{\bar{d}}
\def\e{\epsilon}
\def\wb{\bar{w}}
\def\lb{\bar{\l}}
\def\Jb{\bar{J}}
\def\Nb{\bar{N}}
\def\Zb{\bar{Z}}
\def\pab{\bar{\pa}}

\def\Cb{\bar{C}}

\def\At{\tilde{A}}
\def\Bt{\tilde{B}}
\def\Ct{\tilde{C}}
\def\Dt{\tilde{D}}
\def\Et{\tilde{E}}
\def\Ft{\tilde{F}}
\def\Gt{\tilde{G}}
\def\Ht{\tilde{H}}
\def\Kt{\tilde{K}}
\def\Mt{\tilde{M}}
\def\Nt{\tilde{N}}
\def\Rt{\tilde{R}}
\def\at{\tilde{a}}
\def\bt{\tilde{b}}
\def\ct{\tilde{c}}
\def\dt{\tilde{d}}
\def\et{\tilde{e}}
\def\ft{\tilde{f}}
\def\htil{\tilde{h}}
\def\gt{\tilde{g}}
\def\nt{\tilde{n}}
\def\mut{\tilde{\mu}}
\def\nut{\tilde{\nu}}
\def\pht{\tilde{\f}}
\def\vft{\tilde{\vf}}

\def\rht{\tilde{\rho}}

\def\asth{\hat{*}}
\def\phh{\hat{\phi}}

\def\bA{{\bf A}}

\def\ola{\overleftarrow}
\def\ora{\overrightarrow}
\def\alt{\tilde{\a}}

\def\eh{\hat{e}}
\def\eph{\hat{\e}}
\def\ph{\hat{p}}
\def\alh{\hat{\a}}
\def\beh{\hat{\b}}
\def\gah{\hat{\g}}
\def\Fh{\hat{F}}
\def\muh{\hat{\m}}
\def\nuh{\hat{\n}}
\def\thh{\hat{\th}}
\def\rhh{\hat{\r}}
\def\dh{\hat{d}}
\def\ih{\hat{i}}
\def\jh{\hat{j}}
\def\hh{\hat{h}}
\def\nh{\hat{n}}
\def\gh{\hat{g}}
\def\kh{\hat{k}}
\def\deh{\hat{\d}}
\def\wh{\hat{w}}
\def\lah{\hat{\l}}
\def\Ah{\hat{A}}
\def\Kh{\hat{K}}
\def\Nh{\hat{N}}
\def\Rh{\hat{R}}
\def\Ch{\hat{C}}
\def\Omh{\hat{\Omega}}

\def\xh{\hat{x}}

\def\ps{\rlap{\, /}\;\,p }
\def\ks{\rlap{\, /}\;\,k }

\def\gym{g_{YM}}

\def\adot{\dot{a}}
\def\bdot{\dot{b}}
\def\bpa{\bar{\pa}}

\def\pr{\prime}
\def\ssk{\medskip}
\def\clb{\color{blue}}
\def\clr{\color{red}}
\def\clg{\color{green}}

\begin{document}

\overfullrule=0pt
\parskip=2pt
\parindent=12pt
\headheight=0in \headsep=0in \topmargin=0in
\oddsidemargin=0in

\vspace{ -3cm}
\thispagestyle{empty}
%\vspace{1cm}
%\begin{flushright}
%Preprint DFPD 01/TH/\\
%hep-th/
%\end{flushright}

 \vspace{0.1cm}

\setcounter{equation}{0}
\setcounter{footnote}{0}
\setcounter{section}{0}

\begin{center}

{\Large\bf  Reduction of gravity-matter and dS gravity to hypersurface}

\vskip 0.8cm

 \vspace{.5cm}

\vspace{0.5cm}
I. Y. Park
\\

\vspace{0.3cm}

\vspace{0.3cm}
{\it Department of Applied Mathematics,
Philander Smith College %\footnote{Home institute}
                               \\
Little Rock, AR 72223, USA \\
inyongpark05@gmail.com
}

\end{center}

 \vspace{0.1cm}

\begin{abstract}

\ni The quantization scheme based on reduction of the physical states \cite{Park:2014tia} is extended to two gravity-matter systems and pure dS gravity. For the gravity-matter systems we focus on quantization in a flat background for simplicity, and renormalizability is established through gauge-fixing of matter degrees of freedom. Quantization of pure dS gravity has several new novel features. It is noted that the infrared divergence does not arise in the present scheme of quantization. The lapse function constraint plays a crucial role.

\end{abstract}

\newpage

%%%%%%%%%%%%%%%%%%%%%%%%%%%%%%%%%%%%%
%%%%%%%%%%%%%%%%%%%%%%%%%%%%%%%%%%%%%
\section{Introduction}
%%%%%%%%%%%%%%%%%%%%%%%%%%%%%%%%%%%%%
%%%%%%%%%%%%%%%%%%%%%%%%%%%%%%%%%%%%%

Quantization of 4D Einstein gravity has a long history and has been tackled from various different directions \cite{'tHooft:1974bx,Deser:1974cz,Goroff:1985th,Stelle:1976gc,Antoniadis:1986tu,
Weinberg3,Reuter:1996cp,Odintsov:1990qq,Carlip:2015asa,Thiemann:2007zz,Ambjorn:2012jv,Calcagni:2012hb}. Needless to say, the unrenormalizability has been a major obstacle to progress requiring direct computation in quantum theory of general relativity.
The large amount of gauge symmetry manifest in the ADM formalism of 4D Einstein gravity has recently been exploited to 
lead to explicit and quantitative reduction of the physical states of the theory. That in turn has led to perturbative renormalizability of the theory \cite{Park:2014tia} through a field redefinition of the metric. {(As we will comment below, some aspects of the quantization scheme of \cite{Park:2014tia} have certain similarities with \cite{Modesto:2005sj,Oeckl:2003vu}.) The renormalizability is valid only when the external states satisfy certain physical state conditions (thus become three-dimensional and measure-zero as compared with the offshell states) and thus is not in conflict with the offshell nonrenormalizability established in the past. In other words, the renormalizability established in \cite{Park:2014tia} and subsequent related works is renormalizability of the S-matrix but not that of the offshell Green's functions.} In this work we examine whether or not the same conclusion can be drawn for gravity-matter systems and pure (i.e., without matter) dS gravity.\footnote{The present scheme should be applicable to an AdS spacetime as well at the technical level. However, more care would be needed since an AdS spacetime has well-known issues such as lack of the asymptotic states in the usual sense.} We show that these systems do reduce ``holographically" and are subject to the quantization scheme of \cite{Park:2014tia}.

{Various reductions in the true degrees of freedom were reported in the past in \cite{York:1972sj,Fischer:1996qg,Gay-Balmaz:2014ena,Gomes:2010fh} (see also \cite{yao}). All these works employed the usual 3+1 splitting with the genuine time coordinate separated out. (In contrast, one of the spatial directions was separated out in our works.) Notably, it was observed in \cite{York:1972sj} that the spacelike hypersurface specified up to a conformal factor can be taken as the true degrees of freedom. 
In \cite{Fischer:1996qg}, it was shown in the Hamiltonian formalism that the reduced Hamiltonian was given by the volume of a cerain hypersurface after Hamiltonian reduction was carried out on a class of 4D manifolds with certain topological restrictions. As for the present and related works \cite{Park:2014tia,Park:2014qoa,Park:2015qxa,Park:2014noa} the quantization has been addressed as well: after the foliation-based reduction of the true degrees of freedom is observed, the reduction is shown to lead to quantization. 
For quantization in the Hamiltonian formulation, one must deal with the constraints, the so-called spatial diffeomorphism and Hamiltonian constraints, and how to deal with the diffeomorphism constraint has been one of the major obstacles in the gravity quantization.   
In our works in which both the Hamilatonian and Lagrangian formalisms were adopted, these constraints were called the shift vector and lapse function constraints, respectively. It turned out that the shift vector constraint could be explicitly solved by using an mathematical identity on a commutator of a Lie derivative and covariant derivative as explained in \cite{Park:2014qoa}.\footnote{The account in \cite{Park:2014qoa} was in terms of the component notation. The corresponding ``coordinate-free" version has recently been given in \cite{Park:2016zgt}.} The implication of the solution of the shift vector constraint has been brought out in \cite{Park:2014qoa,Park:2015qxa} by foliation theory. Throughout these works the strategy for reduction has been removal of all of the unphysical degrees of freedom from the external states. The key observation for the reduction was the fact that the residual 3D gauge symmetry - whose detailed analysis is given in \cite{Park:2014noa} - can be employed to gauge away the non-dynamical fields such as the lapse function and shift vector even after the standard bulk gauge-fixing such as de Donder gauge. 

For the renormalization, only the physical states are chosen to be the external legs (namely, we narrow down to the renormalization of the S-matrix), which corresponds to setting the fields in the 1PI effective action to be three-dimensional in a certain sense to be explained below (see also \cite{Park:2015xoa}).
 All of the Riemann tensor terms appearing in the effective action reduce to the well-known expression in terms of the Ricci tensor and metric. In other words, the whole effective action can now be expressed in terms of the 3D Ricci tensor and metric. One can then introduce a metric field redefinition \cite{'tHooft:1973us} to absorb the counterterms and thereby establish renormalizability of the original 4D action.\footnote{{It may be helpful to recall the reason for the offshell nonrenormalizability \cite{'tHooft:1974bx}. (This will in turn provide the rationale for the renormalizability for the physical states.) Let us take the Einstein gravity.  The counterterms such as $R^2, R_{\m\n}R^{\m\n}, R_{\m\n\r\s}R^{\m\n\r\s}$ appear in the one-loop calculation, and unlike the first two, the third type cannot be absorbed by a metric field redefinition. However, it turns out that the combination $R_{\m\n\r\s}R^{\m\n\r\s}-4R_{\m\n}R^{\m\n}+R^2$ is a topological invariant and thus one can trade, for the equation of motion, $R_{\m\n\r\s}R^{\m\n\r\s}$ for the other types:
\bea
R_{\m\n\r\s}R^{\m\n\r\s}=4R_{\m\n}R^{\m\n}-R^2
\eea 
However, such an identity is not available to absorb the counterterms appearing in two- and higher- loop orders. For instance, the counterterm of the form, $R^{\m\n}{}_{\r\s}R^{\r\s}{}_{\a\b}R^{\a\b}{}_{\m\n}$, appears in the two-loop but cannot be absorbed by a metric field redefinition, and thus declared was the offshell nonrenormalizability. The observation in \cite{Park:2014tia} and the subsequent works is that the Riemann tensor itself becomes expressible in terms of the Ricci tensor and metric once only the physical states are consider on the external lines of the Feynman diagrams essentially because the physical states have support on a certain 3D hypersurface.    
}}
Therefore, the gravity renormalizability is more complicated than the nongravitational cases in which only shifts in the parameters are required.

It is well known that, once gravity is coupled with matter, the divergence becomes worse - this poses a challenge to the quantization scheme proposed in \cite{Park:2014tia}.
First of all, there is the question of whether or not a gravity-matter system would reduce in a way similar to the pure gravity case. As a matter of fact, it was initially expected that a gravity-matter system might not reduce, or if it reduces, would reduce in a more complicated way. Secondly, if it does reduce, what would be the steps required to achieve the reduction? Thirdly, would there be a mathematical picture similar to the one in the pure Einstein gravity case? In this work, we show that the answer to the first question is affirmative, and work out the answer to the second question. We also address the third question.

Whereas the reduction of a gravity-matter system is certainly challenging, it would not be entirely natural if the (physical) matter fields - which live ``in'' the spacetime background governed by gravity - do not reduce while their ``habitat" - the spacetime - reduces.
Below we consider two gravity-matter systems, a gravity-(non-self-interacting) scalar system and an Einstein-Maxwell theory, and show that their physical states do reduce. Afterwards we analyze a pure dS gravity system and reach the same conclusion.

As compared to the previous works of \cite{Park:2014tia,Park:2014qoa,Park:2014noa} we employ below more general tactics of achieving reduction by focusing
on the physical meanings of the lapse function constraint. 
One of the key elements is that the lapse function constraint - which is a {\em constraint} - becomes identical to the ``Hamiltonian" itself after the shift vector constraint is solved (which was one of the key observations in \cite{Park:2014tia}) and the gauge-fixing is explicitly enforced. In other words, the lapse constraint generates the ``time"-translation and at the same time it constrains the physical state to be invariant under this translation. What had not been noticed (to our knowledge) in the past was that the shift vector constraint can be explicitly solved \cite{Park:2014tia,Park:2014qoa} in the Lagrangian formulism and omitted from the original Hamiltonian; the resulting Hamiltonian then becomes identical to the lapse function constraint.

As will be analyzed below, the lapse function constraint can be identified with the Halmiltonian density ${\cal H}$ once the shift vector is gauged away: the lapse function constraint takes the form of ${\cal H}|phys>=0$. We take this constraint as to imply that the {\em bulk} degrees of freedom should be in a ``vacuum"\footnote{The ``vacuum" collectively represents the states only with the boundary excitations. In \cite{Higuchi:1991tk} a similar observation - of which we became aware after this work was completed - was made at a linear level. (However, the vacuum in \cite{Higuchi:1991tk} meant the usual vacuum, the state without any excitation.)}. (What precisely we  mean by the vacuum state for the dS case will be specified in the main body.) Even though the bulk degrees of freedom are in a vacuum, it should be perfectly possible to have excitation of ``boundary'' fluctuating degrees of freedom\footnote{The ``boundary" degrees of freedom mean the degrees of freedom that are typically associated with a hypersurface in an "asymptotic" region. As we will see in the section where we analyze the dS case, the physical state constraint implies vanishing of $\w$, one of the quantum numbers.} (the suppression of the bulk degrees of freedom and excitation of the boundary degrees of freedom may be viewed to some extent as a reinterpretation of the Wheeler-DeWitt equation), the ``zero-modes." (See, e.g., \cite{Freidel:2016bxd} and \cite{Donnelly:2016auv} for related interesting recent development.)
In the case of the dS gravity for which the radial direction is split out, the ``boundary'' may be taken as a generic hypersurface at fixed radial coordinate which will be taken at the cosmological horizon at the end.
The upshot is that the lapse function constraint leads to reduction of the physical states and one may focus on the three-dimensional description of the original 4D dynamics, namely the dynamics of the ``boundary" degrees of freedom. This reduction is essentially what makes the quantization possible. The 3D theory is not a genuine three-dimensional gravity as heavily emphasized in \cite{Park:2014tia}: it is a description of the 4D physics through the 3D window. The technical side of the implementation of the reduction will be presented in the main body.
%%%%%%%%%%%%%%%%%%%%%%%%%%%%%%%%%%%%
Once reduction is established one can proceed with the standard perturbation procedure 
as in \cite{Park:2014noa,Park:2015ota,Park:2015xoa}.

As we will see, the materialization of the matter degrees of freedom into those of the metric seems crucial in establishing renormalizability in the matter-gravity systems. (The materialization is nominal to some extent though as will be discussed.)
One of the systems that we consider below is a free (i.e., non-self-interacting) scalar coupled with gravity. The scalar can be gauge-fixed away and absorbed into the metric (details will be given in the main body), thereby guaranteeing renormalizability by the earlier work of \cite{Park:2014tia}. 
Renormalizability of the Einstein-Maxwell system can be similarly established.

The de Sitter case unveils several new features of the present quantization scheme.
We contemplate the meaning of the physical state condition and how it leads to reduction. The quantization around a flat background is simple enough so that one does not need to confront this subtle issue (even in the presence of matter); the issue emerged in the Schwarzschild case to a moderate extent \cite{Park:2015xoa}. In the present work we introduce an interpretation of the physical state condition applicable to all of these previous cases as well.  
In particular, the $r$-dependence ($r$ denotes the radial coordinate) of the coefficients of annihilation and creation operators of the mode expansion is maintained initially, which is eventually taken to the boundary value.  
Another feature of the dS case is that, unlike the flat case, the partially 4D covariant approach \cite{Park:2014noa} may not be applicable (at least without additional consideration and/or modification of the scheme; we will have more in the conclusion) because of the well-known infrared divergence \cite{Ford:1977in,Allen:1985ux,Allen:1987tz}. We adopt the explicitly 3D approach along the line of \cite{Park:2014noa} and show that therein the infrared divergence no longer arises.

\vspace{.3in}
The rest of the paper is organized as follows.
\vspace{.1in}

In section 2, we review the quantization procedure of the pure Einstein gravity in the ADM formalism and give an overview of the reduction scheme by focusing on the double roles of the lapse function constraint: one as the physical state constraint and the other as the generator of the ``time"-translation.
We also comment on the relationship between the lapse function constraint and the abelian Lie algebra associated with the totally geodesic foliation that played an important role in the mathematical picture presented in \cite{Park:2014qoa} and \cite{Park:2015qxa}. 
In section 3, we consider two gravity-matter systems and their quantization around a flat spacetime.\footnote{Interestingly, it appears that a primitive form of a hypersurface reduction was observed long ago in \cite{yao} in the context of flat-space electrodynamics.} We start with a minimally-coupled scalar system and show that once the synchronous gauge is chosen the system reduces and becomes amenable to quantization. Renormalizability is established by gauging away the scalar.
The analysis of the Einstein-Maxwell system parallels the scalar case; it is shown that this system becomes renormalizable by the same mechanism.
In section 4 we consider pure dS gravity in the static coordinate.  
Application of the quantization to a dS background has several additional novel features as compared with the Schwarzschild case \cite{Park:2015xoa}, which itself required more care than the flat case. One of the subtleties is with regards to the location of the boundary. Another subtlety is due to the well-known infrared divergence. As we will see, the infrared divergence does not arise in our 3D setup due to the fact that the sum over the modes associated with the radial direction - which causes the divergence in the conventional setup - is absent. We point out that the present scheme offers additional insights to the observations made in \cite{Allen:1985ux} on reduction of the isometry symmetry (i.e., reduction of the full de Sitter symmetry to its maximal subgroup $E_3$, the isometry of a 3D Euclidean space).
We conclude with summary and future directions in section 5. Appendix A has our conventions and Appendix B has details of the scalar gauge-fixing procedure.

%%%%%%%%%%%%%%%%%%%%%%%%%%%%%%%%%%%%%
%%%%%%%%%%%%%%%%%%%%%%%%%%%%%%%%%%%%%
\section{Review of reduction and quantization}
%%%%%%%%%%%%%%%%%%%%%%%%%%%%%%%%%%%%%
%%%%%%%%%%%%%%%%%%%%%%%%%%%%%%%%%%%%%

In this section we illustrate the quantization scheme proposed in \cite{Park:2014tia} by taking the case of pure Einstein gravity. {The approach of \cite{Park:2014tia} has elements from both the canonical
and covariant approaches: it uses 3+1 splitting of the spacetime and starts
with the ADM Hamiltonian quantization. However, instead of remaining
entirely in the Hamiltonian quantization, a background-specific quantization is considered in the combined Hamiltonian and Lagrangian formalisms.}
We employ the setup of the operator quantization\footnote{{Once the physical states are uncovered, one may also consider the path integral quantization. Again one of the key ingredients for the (restricted) renormalizability is to keep the external states physical.}} in order to uncover the physical states and focus on the dual roles of the lapse function constraint: the physical state constraint as well as the ``usual" translational generator (once the synchronous gauge is taken) along the direction separated out. We will see that it is these dual roles that bring along the reduction.

The gist of the analysis is that the lapse function constraint is at the heart of the reduction:
\bea
\Big[\cR-K^2+K_{mn}K^{mn}\Big]|phys>=0 
\la{NnconHq}
\eea
(See \rf{K4defqq} below for definitions of $\cR, K$ and $K_{mn}$.)
We take this condition as to imply that the {\em bulk} states must be in the vacuum. Nevertheless, the ``boundary" degrees of freedom associated with a hypersurface in a far region may be excited. 
In other words, the operators in the asymptotic boundary region will remain invariant under translation along the direction separated out.\footnote{This is based on the mathematically well known fact that $\infty$  remains as the ``same" $\infty$ even after a finite number is added to it; $\infty$ is not a fixed number but a ``state" of increasing numbers. \la{fninfty}} For a system with rotational symmetry,
 the boundary may initially be taken as a generic hypersurface at a fixed value of $r$ for convenience. One eventually sets $r$ to $r=r_{c.h.}$ (where $r_{c.h.}$ denotes the cosmological event horizon) for the de Sitter case as we will discuss (for the Schwarzschild case considered in \cite{Park:2015xoa} the boundary location was taken at $r=\infty$).

\subsection{Pure Einstein gravity in ADM formalism}

Consider the Einstein-Hilbert action
%%%
\bea
S\equiv \int\sqrt{-{g}}\;{R}
\eea
%%%
with 
%%%
\bea
x^\m\equiv (y^m,x^3)
\eea
%%%
where $\m=0,..,3$ and $m=0,1,2$. (See Appendix A for the conventions.)
Let us parameterize the 4D metric according to 
%%%
\bea
g_{\m\n}=\left(
\begin{array}{cc}
h_{mn} & N_{ m} \\
&\\
N_{ n} &  n^2+h^{mn}N_{m} N_{ n} 
\end{array}
\right)\quad,\quad
g^{\m\n}=\left(
\begin{array}{cc}
h^{mn}+\fr1{n^2}N^m N^n & -\fr1{n^2}N^{ m} \\
&\\
-\fr1{n^2}N^{ n} &  \fr1{n^2} 
\end{array}
\right)
\eea
%%%
One gets (see, e.g., \cite{Arnowitt:1962hi,Poisson} for a review\footnote{The equivalence between the usual formulation and ADM formulation
	of general relativity was questioned in \cite{Frolov:2008sn}.})
\bea
S = \int d^4 x\;n\sqrt{-\g} \Big[\cR+K^2-K_{mn}K^{mn}+2\nabla_\a(n^\b \nabla_\b n^\a-n^\a \nabla_\b n^\b)\Big]
\eea
with
%%%
\be
K_{mn}=\fr1{2n}\left(\mathscr{L}_{{3}} h_{mn}-{\nabla}_m N_{n}
         -{\nabla}_n N_{ m} \right),\qquad K=h^{mn}K_{mn}.
\la{K4defqq}
\ee
%%%
where $\mathscr{L}_{{3}}$ denotes the Lie derivative along the vector field $\pa_{x^3}$ and $\N_m$ is the 3D covariant derivative.
The last term, $2\nabla_\a(n^\b \nabla_\b n^\a-n^\a \nabla_\b n^\b)$ (where $n^\a$ denotes the unit normal to the boundary), is the surface term and will be set aside for the quantization in the flat background below.
The ``spatial" components of the bulk de Donder gauge $g^{\r\s}\G^\m_{\r\s}=0$ read, in the ADM fields \cite{Smarr:1978dia},
%%%
\bea
&& (\pa_{x^3}-N^n \pa_n)N^m=n^2(h^{mn}\pa_n \ln n-h^{pq}\G^m_{pq})
\la{ADMddq}
\eea
%%%
One way of achieving reduction is to fix, by using the gauge symmetry generated by the $x_3$ component of the diffeomorphism parameter that has a residual 3D coordinate dependence \cite{Park:2014noa}:
\bea
n=n_0
\eea
where $n_0$ denotes the background value and $n_0=1$ for a flat background. (For the gravity-matter cases considered below, $n$ will not directly be gauge-fixed; one of the components of the matter fields will be gauge-fixed instead.) The trace piece of the metric is also gauge-fixed.
As discussed in \cite{Park:2015xoa}, the resulting equation $K=0$ - which can be viewed as the $x^3$-component of the nonlinear de Donder gauge - can be interpreted as the constraint associated with this fixing.
We also adopt the synchronous-type gauge-fixing of the residual 3D gauge symmetry:
\bea
 N_m=0
\eea
(For a flat background this gauge can definitely be chosen. More generally, we consider only the backgrounds that are compatible with this gauge\footnote{The $N_m=0$ gauge can always be chosen according to \cite{Landau}.} and have relatively simple foliation. Schwarzschild and dS backgrounds are among the examples.)
The induced shift vector constraint,
\bea
\N^n(-K_{mn}+K h_{mn})=0,
\eea
is automatically satisfied \cite{Park:2014tia} with the gauge-fixing above, $n_0=1$.
We impose the lapse function constraint (i.e., the field equation of the lapse field) as the physical state condition
%%%
\bea
\Big[\cR-K^2+K_{mn}K^{mn}\Big]|phys>=0 
\la{NnconH}
\eea
One can go to the Hamiltonian formulation by the usual Legendre transformation, to which we now turn. 

Before imposing the gauge-fixing, the bulk part of the ``Hamiltonian" of $x^3$-evolution takes
%%%
\bea
\!\!\!\!\!\!  H=\int d^3y\left[ n( -h)^{-1/2}({ -}\pi^{mn}\pi_{mn} { +} \fr12 \pi^2)
-n (-h)^{1/2}R^\3-2N_m (-h)^{1/2}\N_n[(-h)^{-1/2}\pi^{mn}]  \right]
\nn\\
\eea
%%%
where $\pi^{mn}$ denotes the momentum field,
\bea
\pi^{mn}=\sqrt{{-}h}\;({ -}K^{mn} { +}K h^{mn})
\eea
%%%
Let us express the Hamiltonian in terms of the lapse and shift constraints.
Omitting the surface terms, the Hamiltonian density is given by
\bea
{\mathscr H} &=& \sqrt{-h}\, n(K^2-K_{mn}K^{mn}-\cR)+2\sqrt{-h}\, N_m\N_n(K^{mn}-K h^{mn})\nn\\
  &=& \sqrt{-h}\,\Big[-n {\cal C}_0-{2}N^{m}{\cal C}_{0m} \Big]
\eea
where
\bea
{\cal C}_0\equiv \cR-K^2+K_{mn}K^{mn} \nn\\
{\cal C}_{0m} \equiv \N^n(-K_{mn}+K h_{mn})
\eea
In the Hamiltonian formalism these are called the Hamiltonian and momentum constraints respectively; we have been informally calling them the lapse function and shift vector constraints in the context of the Lagrangian analysis. 
Let us now specialize to quantization in a flat background.
As discussed in details in \cite{Park:2015xoa}, the gauge-fixing of the trace piece of the metric leads to the following constraint:
\bea
K=0
\eea 
This is consistent with the $x^3$ component of the nonlinear de Donder gauge that we have discussed above (see the appendix for the full set of the nonlinear de Donder gauge). 
The curved space generalization was discussed in \cite{Park:2015xoa} and will be briefly reviewed in section 4.
The following gauge is suitable for quantization in a flat background
\bea
n=1 \quad,\quad N_m=0  \la{gf}
\eea 
With $n=1$, the shift vector constraint is automatically satisfied and the Hamiltonian density takes
\bea
{\mathscr H}_{g.f.} &=& -\sqrt{-h}\: {\cal C}=-\sqrt{-h}\;(\cR-K^2+K_{mn}K^{mn})
\eea
where ${\mathscr H}_{g.f.}$ denotes the gauge-fixed Hamiltonian density.
%Let us split the Hamiltonian $H_{g.f.}=\int d^3y {\mathscr H}_{g.f.}$ into a free and an %interacting parts:
%\bea
%H_{g.f.} &=& H_{0g.f.}+\tilde{H}_{g.f.}
%\eea 
The lapse function constraint \rf{NnconH} with the gauge-fixing \rf{gf} implies
\bea
H_{g.f.}|phys>=0  \la{Hconstr}
\eea 
where $H_{g.f.}$ denotes the gauge-fixed Hamiltonian.
%This is in the Heisenberg picture. 
It is in the full nonlinear sense. 
Note the dual roles of the gauge-fixed Hamiltonian: it is the operator that governs the $x^3$-evolution, and acts as a constraint at the same time. 
As previously stated, this should indicate that the bulk state must be in a vacuum. (We will come back to this in the dS analysis in section 4.) It is this reduction that allows a description of the 4D physics through the 3D window. In section 3, we will repeat the analogous steps for the matter and dS systems. With the reduction established one can follow the perturbative renormalization procedure such as those in \cite{Park:2014noa} and \cite{Park:2015xoa}.

%%%%
\begin{figure}
\centerline{
\begin{minipage}[b]{8cm}
             \epsfxsize=8cm
              \epsfbox{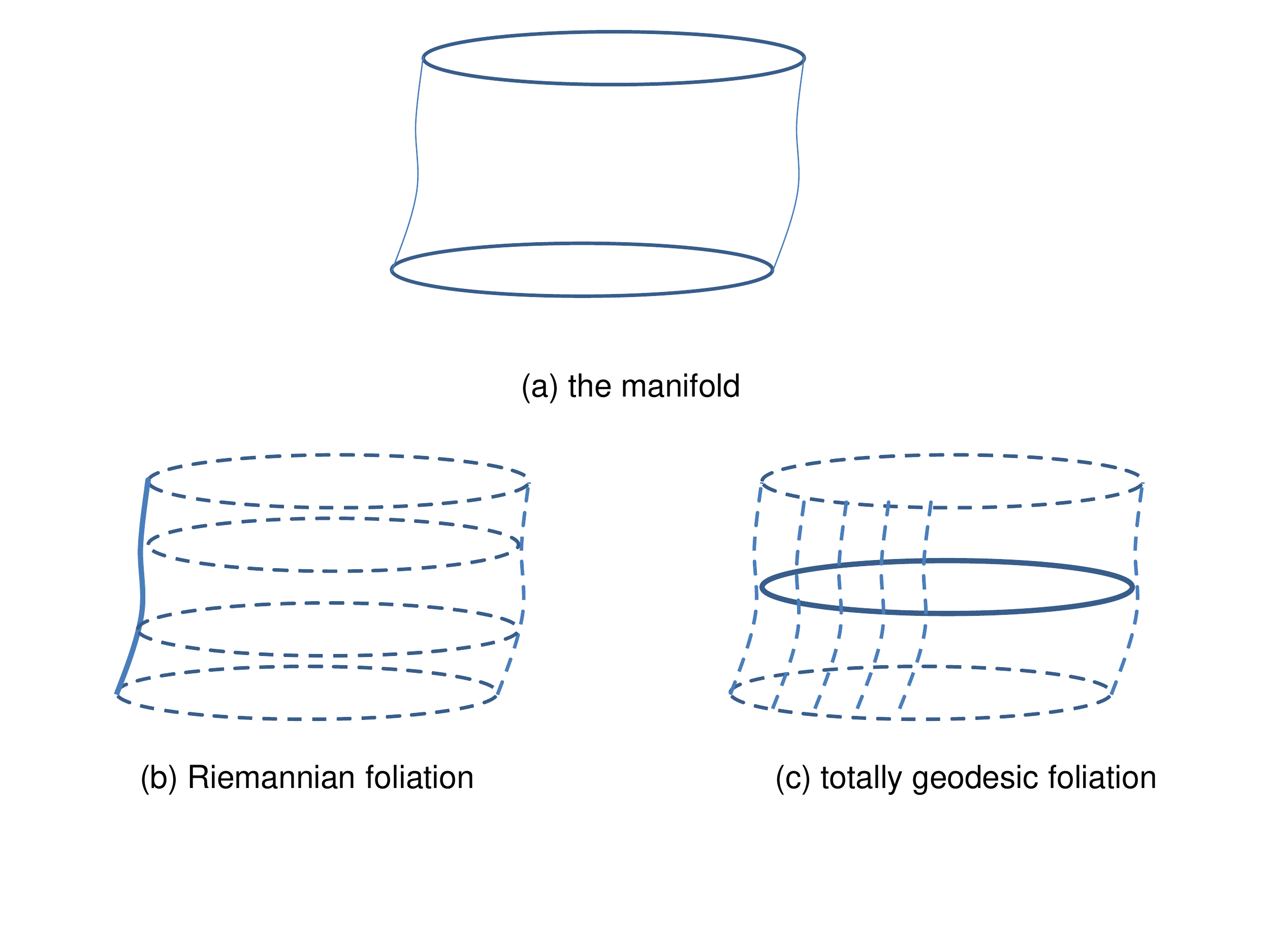}
      \end{minipage}
      }
\caption{duality in foliations: the solid lines in (b) and (c) represent the base (i.e., the space of the leaves) in each case}
\label{fig1}
\end{figure}

{
\subsection{renormalization via metric field redefinition}

As shown long ago, gravity renormalization involves a metric field redefinition in general \cite{'tHooft:1973us}\cite{'tHooft:1974bx}.
The Einstein gravity is one-loop renormalizable due to the following offshell identity:
\bea
R_{\m\n\r\s}R^{\m\n\r\s}-4R_{\m\n}R^{\m\n}+R^2=\mbox{total derivative} \la{Riemannsqid}
\eea
Because of this it is not necessary to employ the reduction of the physical states at (and only at) one-loop. 
Let us consider the one-loop diagram in the graviton sector with the graviton loop by employing the traceless propagator. The importance of the traceless propagator has been analyzed in detail in \cite{Park:2015ota}\footnote{As a matter of fact, the necessity of the traceless gauge-fixing was noticed in \cite{Ortin} (see ch. 3) much earlier as we have become aware of reecntly.}: in the literature the covariance of the terms in the 1PI action is {\em presumed} and their coefficients are subsequently determined. The analysis in \cite{Park:2015ota} has revealed that the presence of the trace piece makes this step invalid because the trace mode destroys the covariance. The Einstein action with the de Donder gauge-fixing term can be expanded to the relevant order to yield
\bea
\cL&=& -\fr12 {\pa}_\g h^{\a\b}{\pa}^\g h_{\a\b}+\fr14 {\pa}_\g h^{\a}_\a {\pa}^\g h^{\b}_\b  
 + \cL_{V_1}+\cL_{V_{2}}   \la{eawv}
\eea
where 
\bea
\hspace{-.3in}\cL_{V_1} &=&   \Big(2\eta^{\b\b'}{\G}^{\a' \g\a}- \eta^{\a\b}{\G}^{\a' \g\b'}\Big)\pa_\g h_{\a\b}\, h_{\a'\b'}  
+ \Big[\fr12(\eta^{\a\a'}\eta^{\b\b'}\vf^{\g\g'}+\eta^{\b\b'}\eta^{\g\g'}\vf^{\a\a'}
+\eta^{\a\a'}\eta^{\g\g'}\vf^{\b\b'})\nn\\
\hspace{-.2in}&& -\fr14 \vf\, \eta^{\a\a'}\eta^{\b\b'}\eta^{\g\g'}-\fr12 \eta^{\g\g'}\eta^{\a'\b'}\vf^{\a\b} 
+\fr14 (-\vf^{\g\g'}+\fr12 \vf \eta^{\g\g'})\eta^{\a\b}\eta^{\a'\b'}
\Big] \pa_\g h_{\a\b}\, \pa_{\g'}h_{\a'\b'} 
\eea
\bea
\cL_{V_{2}} = \sqrt{-g}\Big( h_{\a\b}h_{\g\d}R^{\a\g\b\d}-h_{\a\b}h^{\b}{}_\g R^{\k\a\g}{}_{\k}
+h^{\a}{}_{\a}h_{\b\g}R^{\b\g}-\fr12 h^{\a\b}h_{\a\b}R
+\fr14  h^{\a}_\a  h^{\b}_\b R \Big)  \la{gver}  \nn\\
\eea
One can show that the counterterms associated with $\cL_{V_{2}} $ are given by\footnote{{The conventional way of applying the background field method with the traceful propagator leads to non-covariant forms of the counterterms, a problem actually known in the literature. (See, e.g., ch. 3 of \cite{Buchbinder}.)}}
\bea
\D \cL &=& \fr{\G(\e)}{(4\pi)^2}\Big[-\fr54R^2-\fr34 R_{\m\n\r\s}R^{\m\n\r\s} +\fr32R_{\m\n}R^{\m\n}\Big]   \nn\\
&=& -\fr12\fr{\G(\e)}{(4\pi)^2}\Big[-3R_{\m\n}R^{\m\n}-R^2\Big]  \la{tgravi}
\eea
where in the third equality the identity \rf{Riemannsqid} has been used. Including the $\cL_{V_{1}}$ and the ghost-loop contributions, the total one-loop counterterms are given by \cite{Park:2015ota}
\bea
\D \cL_{1loop} 
&=& -\fr12\fr{\G(\e)}{(4\pi)^2}\Big[-\fr{41}{60}R_{\m\n}R^{\m\n}{ -\fr{27}{40}}R^2\Big]\la{totalctr}
\eea
These counterterms can be absorbed by the following metric redefinition:
\bea
g_{\m\n}\ra g_{\m\n}+c_1 \k^2 g_{\m\n}R+c_2 \k^2R_{\m\n};
\eea
with
\bea
c_1=\fr{61}{120}\fr{\G(\e)}{(4\pi)^2} \quad,\quad  c_2= -\fr{41}{120}\fr{\G(\e)}{(4\pi)^2} 
\eea	
Note that this would not have been possible had it not been for the identity \rf{Riemannsqid} by which the $R_{\m\n\r\s}R^{\m\n\r\s}$-term is converted into the Ricci tensor and scalar terms. As will be reviewed in section 4.1, this analysis has been extended to a curved background \cite{Park:2015xoa} where the application to the higher loops has also been outlined.

}

\subsection{connection with mathematical approach}

The mathematical approach of \cite{Park:2014qoa}\cite{Park:2015qxa} hinges on the duality between Riemannian foliation and totally geodesic foliation (see Fig. \ref{fig1}).
The abelian parallelism of the totally geodesic foliation should be associated with the lapse function constraint in the gauge chosen above.
This can be seen from the fact that gauge-fixing of the lapse function leads to reduction to 3D, which corresponds, in the mathematical picture, to the fact that upon modding out by $U(1)$ the jet bundle yields the 3D connection \cite{Park:2015qxa}. Thus the physical picture associated with the lapse and shift gauge-fixings suggests that the abelian algebra is associated with the gauge symmetry. 
More specifically, with the shift vector gauge-fixing above, the operator $R^{(3)}-K^2+K_{mn}K^{mn}$ should be a representation of the generator of the abelian symmetry
of the totally geodesic foliation.

%%%%%%%%%%%%%%%%%%%%%%%%%%%%%%%%%%%%%
%%%%%%%%%%%%%%%%%%%%%%%%%%%%%%%%%%%%%
\section{Gravity-matter systems}
%%%%%%%%%%%%%%%%%%%%%%%%%%%%%%%%%%%%%
%%%%%%%%%%%%%%%%%%%%%%%%%%%%%%%%%%%%%

In this section we consider quantization of two gravity-matter systems in a flat background.
Since the matter fields live ``in" the gravity background, it seems natural, to some extent, to expect that they should reduce as well. 
The analysis indicates that the reduction does occur and occurs rather generically: the key point remains the same as the pure gravity case just reviewed: the ``Hamiltonian" - including the matter part - generates the $x^3$-translation; once the shift vector is fixed to zero, the Hamiltonian becomes identical to the lapse function constraint. (The shift vector constraint can be explicitly solved \cite{Park:2014qoa} and omitted from the original Hamiltonian.)

In both of the cases, some matter components can potentially come to materializes into additional metric degrees of freedom by gauge-fixing (but see footnote \ref{tfn}): for the scalar case the scalar field may reappear as one of the metric degrees of freedom.\footnote{One may wonder whether the scalar field could be absorbed  by a metric field redefinition instead of being gauge-fixed. 
The field redefinition does not make the scalar non-dynamical as can be seen from the analysis of, e.g., \cite{Jarv:2014hma}. (I thank its authors for making this point clear.)} In the Maxwell-Einstein case, two of the vector field components are converted into the corresponding metric degrees of freedom. 
Let us count the number of the gauge parameters and the corresponding fixings in the gravity-scalar case to insure that they match: the three bulk gauge parameters can be used for the $\m=0,1,2$ components of the de Donder gauge, the three 3D gauge parameters for the shift vector fixing (see the analysis of the residual symmetry in \cite{Park:2014noa}), and lastly one parameter with the 3D coordinate dependence for gauge-fixing of the scalar since the scalar field at this point becomes reduced to 3D.\footnote{One may adopt a slightly different gauge-fixing strategy: for instance the four bulk parameters could be used to gauge away the shift vector and the scalar. Then the system reduces to 3D and therefore the 3D parameters can be used to fix the rest of the fields.}
	The metric-trace fixing can be effectively executed by employing the traceless propagator and the lapse function is fixed automatically from these fixings.\footnote{Because of this the conversion of the matter components into the metric degrees of freedom is nominal to some extent: using part of the residual 3D residual gauge symmetry to explicitly gauge away the lapse or the trace piece of the 3D fluctuation metric is not necessary. \la{tfn}} The Einstein-Maxwell system has similar gauge-fixings and will be discussed below.

Below we first consider a scalar-gravity system and carry out quantization around a flat background. With the (fluctuating part of the) scalar gauge-fixed to zero, the system becomes a ``pure" gravity system (since the background scalar vanishes as well for a flat background) and thus becomes renormalizable. We then consider an Einstein-Maxwell system and repeat the analysis.

\subsection{gravity coupled with scalar}

The action for metric minimally coupled with a scalar $\z$ is (see, e.g., \cite{Parker} for a review of a scalar field in a dS background)
%%%
\bea
S=\int\sqrt{-{g}}\;\Big({R}-\fr12g^{\m\n}\pa_\m \z \pa_\n \z \Big)
\la{unsplit}
\eea
%%%
By using the 3+1 splitting 
%%%
\bea
g^{\m\n}=\left(
\begin{array}{cc}
h^{mn}+\fr1{n^2}N^m N^n & -\fr1{n^2}N^{ m} \\
&\\
-\fr1{n^2}N^{ n} &  \fr1{n^2} 
\end{array}
\right);
\eea
%%%
the action can be re-written
%%%
\bea
S &=& \int d^4 x\;n\sqrt{-h} \Big[\cR+K^2-K_{mn}K^{mn}-2\L\nn\\
&&\hspace{-.5in}-\fr12\Big(h^{mn}+\fr1{n^2}N^m N^n\Big) \pa_{m}\z\pa_{n}\z
-\fr1{2n^2}\pa_3 \z \pa_3 \z+\fr1{n^2}\pa_{m}\z \pa_{3}\z N^m
\Big]  
\eea
%%%
The shift vector field equation now has the additional matter part:
\bea
 \nabla^n(-K_{mn}+Kh_{mn})-\fr{N^n}{n} \pa_m \z \pa_n \z  
+\fr{1}{ n} \pa_m \z \pa_3 \z =0
\eea
%%%
%\bea
%{ -}{\N}_a (K^{ab}-h^{ab} K) + \fr1{n}\pa^b\z \pa_{3}\z =0  \la{Nconq}
%\eea
%%%
The lapse function field equation is
\bea
\cR-K^2+K_{pq}K^{pq}-\fr12 \Big(h^{mn} {-\fr{N^m N^n}{n^2} }\Big)\pa_m \z \pa_n \z   { -}\fr{N^m}{n^2}\pa_3 \z \pa_n\z
+\fr1{2n^2}\pa_3 \z \pa_3\z=0  \nn\\
\la{neom}
\eea
As we will see shortly, this constraint becomes identical to the Hamiltonian upon enforcing $N_a=0$ in the Hamiltonian. 
The system then reduces to 3D and the scalar field can be gauged away by using one of the diffeomorphism parameters (the details can be found in Appendix B):
\bea
 \z=0 \quad \la{asser}
\eea
The shift vector field equation then reduces to the pure gravity case:
\bea
\nabla_p(h^{pm}K-K^{pm}) =0
\eea
and it leads to the determination of the lapse function: $n=n_0$ ($=1$ for the flat case\footnote{See \cite{Park:2014qoa} and \cite{Park:2016zgt} for more details on the fact that $\nabla_p(h^{pm}K-K^{pm})$ implies
$n=n_0$.}) once $N_a=0$ is enforced.

The Hamiltonian is given by
\bea
&& \hspace{1.3in}\mathscr{H} = \pi^{ab}\mathscr{L}_{\pa_3} h_{ab}+\pi_{\z}\mathscr{L}_{\pa_3} \z-\cL\nn\\
&& \!\!\!\!\!\!\!\!\!             = -\sqrt{-h}\;(n\,{\cal C}_0 +{2}N^m {\cal C}_{0m})
            +\sqrt{-h}\;n \fr12\Big(h^{mn}+\fr1{n^2}N^m N^n\Big) \pa_m \z \pa_n \z
  -\fr{\sqrt{-h}}{2n} \pa_3 \z \pa_3 \z     \nn\\ 
%%%%%%%%%%%%%%%%%%%%%%%%%%%%%%%%%%%%%%%%%%%5
&=& -\sqrt{-h}\;n\Big[\cR-K^2+K_{mn}K^{mn}-\fr12\Big(h^{mn}-\fr1{n^2}N^m N^n\Big) \pa_m \z \pa_n \z   \nn\\
&& \hspace{1.5in}+\fr1{2n^2}\pa_3 \z \pa_3 \z-\fr1{n^2}\pa_{m}\z \pa_{3}\z N^m\Big] \nn\\
&& -\sqrt{-h}\;N^m\Big[{2}\nabla(-K_{mn}+Kh_{mn})-\fr{N^n}{n} \pa_m \z \pa_n \z  
+\fr{1}{n} \pa_m \z \pa_3 \z   \Big]    
\eea
Noting that
\bea
{\cal C}_\z &\equiv& \cR-K^2+K_{mn}K^{mn}\\
  &&-\fr12\Big(h^{mn}-\fr1{n^2}N^m N^n\Big) \pa_m \z \pa_n \z   +\fr1{2n^2}\pa_3 \z \pa_3 \z-\fr1{n^2}\pa_{m}\z \pa_{3}\z N^m    \nn
\eea
the equation above can be rewritten as\footnote{
it cannot, however, be written as a form analogous to the pure gravity case,
\bea
 -\sqrt{-h}\;(n\,{\cal C}_\z +{2}N^m {\cal C}_{\z m})
\eea
where
\bea
{\cal C}_{\z m} &\equiv&  \nabla^n(-K_{mn}+Kh_{mn})-\fr{N^n}{2n} \pa_m \z \pa_n \z  
+\fr{1}{2n} \pa_m \z \pa_3 \z
\eea }
\bea
\mathscr{H}=-\sqrt{-h}\;n {\cal C}_\z
 -\sqrt{-h}\;N^m\Big[{2}\nabla(-K_{mn}+Kh_{mn})-\fr{N^n}{n} \pa_m \z \pa_n \z  
+\fr{1}{n} \pa_m \z \pa_3 \z   \Big] \nn\\
\eea
With the shift vector gauge-fixing explicitly enforced in the Hamiltonian the Hamiltonian becomes essentially identical to the lapse function constraint. With this the system comes to admit the 3D description and the renormalization procedure of the S-matrix can be followed.

%%%%%%%%%%%%%%%%%%%%%%%%%%%%%%%%%%%%%
\subsection{Einstein-Maxwell system \la{secEM}}
%%%%%%%%%%%%%%%%%%%%%%%%%%%%%%%%%%%%%

Let us consider
%%%
\bea
S=\int d^4 x \sqrt{-g}\Big[R-\fr14 F_{\m\n}F^{\m\n}\Big]
\eea
%%% 
The 3+1 splitting yields
%%%
\bea
S &=& \int d^4 x\;n\sqrt{-\g} \Big[\cR+K^2-K_{mn}K^{mn}\\
&&-\fr14 F_{mn}F^{mn}
-\fr1{2n^2}(F_{3n}F_{3p}h^{np}-2F_{mn}F_{3p}N^m h^{np}
+F_{mn}F_{pq}N^n N^q h^{mp})
\Big]  \nn
\la{1p3act}
\eea
%%%
where the indices are raised and lowered by the 3D metric $h_{mn}$.
%The second line of \rf{1p3act} is the surface terms; they will be set aside until we come back %to them later.
We fix the $U(1)$ gauge invariance associated with the vector field by
\bea
A_3=0
\eea 
The canonical momentum is
\bea
\Pi^m\equiv \fr{\cL}{\pa (\mathscr{L}_{\pa_3} A_m)}=-\sqrt{-g}\;F^{3m}
\eea 
The lapse function and shift vector constraints are given respectively by 
\bea
{\cal C}_0-\fr14 F_{mn}F_{pq}h^{mp}h^{nq}+\fr1{2 (\sqrt{-h}\;)^2} \Pi^m \Pi^n h_{mn} =0
\eea
and
\bea
-\nabla^k (K_{km}-h_{km}K)-\fr{1}{\sqrt{-h}}F_{mk}\Pi^k=0 \la{scim}
\eea
The Hamiltonian now takes
\bea
&& \hspace{1.3in}\mathscr{H} = \pi^{ab}\mathscr{L}_{\pa_3} h_{ab}+\Pi^{m}\mathscr{L}_{\pa_3} A_m-\cL\nn\\
&& \!\!\!\!\!\!\!\!\!             = -\sqrt{-h}\;(n\,{\cal C}_0 +{2}N^m {\cal C}_{0m})
       -\fr{n}{2\sqrt{-h}}\Pi^m \Pi^n h_{mn} +\Pi^k N^q F_{qk}
       +\fr14 n\sqrt{-h}\; F_{mn}F_{pq}h^{mp}h^{nq}        \nn\\ 
%%%%%%%%%%%%%%%%%%%%%%%%%%%%%%%%%%%%%%%%%%%5
&=& -\sqrt{-h}\;n\Big[\cR-K^2+K_{mn}K^{mn}  +\fr1{2 (\sqrt{-h}\;)^2} \Pi^m \Pi^n h_{mn} -\fr14 F_{mn}F_{pq}h^{mp}h^{nq}\Big] \nn\\
&& -\sqrt{-h}\;N^m\Big[{2}\nabla^n(-K_{mn}+Kh_{mn})-
{\fr{1}{\sqrt{-h}}}F_{mk}\Pi^k\Big]   
\eea
Similarly to the scalar case, it can be written as
\bea
\mathscr{H} =-\sqrt{-h}\;n {\cal C}_A
  -\sqrt{-h}\;N^m\Big[{2}\nabla^n(-K_{mn}+Kh_{mn})-
{\fr{1}{\sqrt{-h}}}F_{mk}\Pi^k\Big]
\eea
where
\bea
{\cal C}_A &\equiv& {\cal C}_0-\fr14 F_{mn}F_{pq}h^{mp}h^{nq}+\fr1{2 (\sqrt{-h}\;)^2} \Pi^m \Pi^n h_{mn}  
\eea
Let us gauge-fix
\bea
N_m=0 
\eea 
by exploiting the 3D diffeomorphism; the Hamiltonian becomes identical to the lapse function constraint. The rest of the gauge-fixing will be discussed in the next subsection where renormalizability is established.

%%%%%%%%%%%%%%%%%%%%%%%%%%%%%%%%%%%%%
\subsection{on renormalizability of S-matrix}
%%%%%%%%%%%%%%%%%%%%%%%%%%%%%%%%%%%%%

The presence of the matter fields makes the reduction procedure more complicated.
It is somewhat ironic, but renormalizability can be most easily established by converting one of the matter degrees of freedom into the metric component. (In retrospect what this means is that it would probably be possible to establish the renormalizability without converting the matter into graviton. However one would need to introduce a quite nontrivial field redefinition. We will have more on this in the conclusion.)

Let us consider the scalar system. As previously stated, the renormalizability is guaranteed with the scalar field gauged away and, thereby, the system rendered entirely ``gravitational." Since one of the gauge parameters should be used, the gauge-fixing of the scalar implies one additional component for the metric: the (fluctuating part of the) scalar field has materialized into this additional component.\footnote{The background part of the scalar, in case there exists such a part, will serve as the background for the dynamic gravitons; we will have more on this in the conclusion.} (See, however, footnote \ref{tfn}.)
 
With the vector field, renormalizability can be established by converting two of the vector components to the metric components. To that end one should consider the curved space version of the Coulomb gauge and explicitly remove two of the four components of the vector field. 
Gauging away of the remaining two components can be performed by using the diffeomorphism parameters.
Let us count: the three bulk gauge parameters for spatial part of de Donder, the three 3D parameters for the shift vector fixing, and finally two 3D parameter for two of the remaining ``3D" vector component. (Again employ the traceless graviton propagator for the metric trace-fixing; the lapse function gets fixed automatically.)

%%%%%%%%%%%%%%%%%%%%%%%%%%%%%%%%%%%%%
\section{Pure dS gravity}
%%%%%%%%%%%%%%%%%%%%%%%%%%%%%%%%%%%%%

Because the de Sitter spacetime is a curved background, things become more complicated once expansion around the background is considered. 
In \cite{Park:2015xoa}, the quantization scheme of \cite{Park:2014tia} was applied to a curved background of a Schwarzschild spacetime with the radial coordinate $r$ chosen to be the reduced direction.
The quantization scheme can be applied to a dS spacetime in the static coordinates with the steps in section 2 carried over with minor modifications entailing the cosmological constant term. 
For instance, the physical state condition gets modified to
\bea
\Big[\cR+K_{mn}K^{mn}-K^2    -2\L\Big]|phys>=0  \la{nconstrq2}
\eea
We illustrate some of the salient points stated in the introduction in the context of dS gravity. In particular, we elaborate on the meaning of the reduction and how the constraint above leads to the ``zero-modes." As will be described in more detail below, reduction should mean that the physical states will be defined in the ``asymptotic region." The meaning of the asymptotic region is clear, e.g., in the Schwarzschild case but this is not the case for the dS.
Once the reduction is established, the perturbative analysis can be set by using the ingredients of \cite{Park:2014noa} and \cite{Park:2015xoa} excepting a few conceptual and technical subtleties pertaining to the infrared divergence and choice of vacuum.

%If one solves the linear-level field equation, the solution will be $r$-dependent as we will %review below. The creation and annihilation operators should be $r$-independent whereas the %coefficients of annihilation and creation operators will be r-dependent. The fields $\Phi$ %will satisfy the linear order field equation since the coefficient do.

One of the differences between the Schwarzschild and dS cases is the infrared divergence in the dS case.
The presence of the infrared divergence was noticed long ago \cite{Ford:1977in}. (Afterwards reduction of the full de Sitter symmetry to its maximal subgroup $E_3$, the isometry of a 3D Euclidean space, was noticed in \cite{Allen:1985ux}. Our scheme provides a natural interpretation of this phenomenon as we will discuss below.)
The infrared divergence does not arise in the present 3D setup.

Not having to worry about infrared divergence, one has two options for the intermediate off-shell analysis in the flat or Schwarzschild case: a 3D or 4D approach. The 3D approach was adopted in \cite{Park:2014noa} and the 4D approach in \cite{Park:2015xoa}. It is not clear, however, if such a 4D off-shell approach would make sense in the present dS setup.
In spite of this difference, though, some of the ingredients used in \cite{Park:2015xoa} will be useful for the dS case as well. This is because the difference between the 3D and 4D approaches sets in only when one considers mode sums in computing the propagator. For the initial formal level the difference is immaterial. 
Let us review the quantization procedure around a Schwarzschild background \cite{Park:2015xoa} and set the stage for the dS analysis that follows.

\subsection{review of Schwarzschild case}

The meaning of the physical state condition, 
\bea
\Big[\cR+K_{mn}K^{mn}-K^2\Big]|phys>=0
\eea
is less subtle compared with the dS case discussed below: they are the states constructed out of the operators at the ``boundary," i.e., an asymptotic region $r\ra \infty$. (At the off-shell level (such as in the path integral) the $r$-coordinate is kept arbitrary.) This is because (the locations of) those operators would be invariant under translation generated by the Hamiltonian of $r$-evolution.

In the partially-4D setup, one can use the off-shell degrees of freedom when computing loops. It is only when carrying out the LSZ reduction procedure that one needs to use the external physical states.
After the trace piece of the metric is gauged away (the importance of taking the traceless propagator was emphasized
in \cite{Park:2015xoa}) the kinetic term is given by (our conventions can be found in the appendix)
\bea
\sqrt{-g}\cL_{kin} = -\fr12\sqrt{-g}\; {\N}_\g \f^{\a\b}{\N}^\g \f_{\a\b}
\la{kin}
\eea
The traceless propagator is given by
\bea
<\f_{\m\n}(x_1)\f_{\r\s}(x_2)>=P_{\m\n\r\s}\,\D(x_1-x_2) 
\eea
where $P_{\m\n\r\s}$ denotes
\bea
P_{\m\n\r\s}=\fr12(g_{0\m\r}g_{0\n\s}+g_{0\m\s}g_{0\n\r}-\fr12 g_{0\m\n}g_{0\r\s})
\eea
and $\D$ satisfies\footnote{Works on $\D(x_1-x_2)$ can be found in \cite{Candelas:1980zt} and more recently in \cite{Paszko:2012zz} and \cite{Garcia-Recio:2013ixa}.}
\bea
\nabla_0^\m \nabla_{0\m} \D(x_1-x_2)=\d^\4(x_1-x_2) \la{sl}
\eea
where $\nabla_{0\m}$ denotes the covariant derivative pertaining to the background $g_{0\m\n}$
and $\d^\4$ includes the metric factor.
One can compute the effective action by following the prescription given, e.g., in chapter 16 of \cite{Weinberg2}, and the Riemann tensor and its contractions will appear. Let us illustrate the reduction at the level of 1PI action by taking $R_{\k_1\k_2\k_3\k_4}$ as an example. First, impose the gauge-fixings and constraints: 
\bea
N_m=0,n=n_0(r),\nabla_m n=0
\eea
where the subscript $0$ denotes the background valued field (i.e., with no fluctuation).
With these the Riemann tensor components take
\bea
R_{mnpq}&=&{\cal R}_{mnpq}+K_{mq}K_{np}-K_{mp}K_{nq} \nn\\
R_{3mnp}&=&-n_0(\nabla_n K_{mp}-\nabla_p K_{mn})\nn\\
R_{m3p3}&=&
-n_0\mathscr{L}_{\pa_{r}}K_{mp}            
+n_0^2 K_{mr}K^r_{p}   \la{rrt}
\eea
One of the things shown in \cite{Park:2015xoa} is that one can replace $K_{mn}$ by its background value:
\bea
K_{mn}=K_{0mn}
\eea
Finally use the following relation
\bea
\cR_{mnrs}&=&\Big(\cR_{mr}-\fr14\cR h_{mr}\Big)h_{ns}
-\Big(\cR_{ms}-\fr14\cR h_{ms}\Big)h_{nr}\nn\\
&&+\Big(\cR_{ns}-\fr14 \cR h_{ns}\Big)h_{ms}
-\Big(\cR_{nr}-\fr14 \cR h_{nr}\Big)h_{ms}
\eea
With $\cR_{mnrs}$ now expressed in terms of the Ricci tensor and scalar it will be possible to absorb the counter-terms by the following metric redefinition almost as in the flat case. Because of the presence of the terms that contain $K_{mn}$-factors, another tensor, $s_{\m\n}$, will present compared with the flat case\footnote{An explicit example of the renormalization procedure of a gravity-matter system in a time-dependent background has recently been worked out in \cite{Park:2016zgt}.}:  
\bea
g_{\m\n}\ra g_{\m\n}+c_1  g_{\m\n}R+c_2 R_{\m\n}+c_3 s_{\m\n}  \la{gredef}
\eea

\subsection{reduction of dS gravity}

Most of the steps in \cite{Park:2015xoa} can be carried over with minor modifications. There are several steps that cannot be directly carried over. For example, the dS case requires dealing with the infrared divergence. Another difference is that
the location of the ``boundary" will be taken at $r=r_{h.c.}$ (where $r_{h.c.}$ denotes the cosmological event horizon) for dS instead of at $r=\infty$ for the Schwarzschild case.

Unlike the flat or Schwarzschild case, 
the partially-4D covariant approach \cite{Park:2015ota} doesn't seem applicable, at least in the manner proposed in \cite{Park:2015ota} and \cite{Park:2015xoa}, due to the infrared divergence in the 4D setup. Therefore we employ the 3D approach \cite{Park:2014noa} and the infrared divergence does not arise in this approach. The reason, as we will see shortly, is that with the reduction the sum over the modes of the reduced direction - which caused the divergence in the conventional analysis - becomes irrelevant.

The action,
%%%
\bea
S=\int d^4 x \sqrt{-g}\Big[R-2\L\Big],
\eea
%%% 
admits the following dS solution in the static coordinate
\bea
ds^2=-\Big(1-{\l^2}{r^2}\Big)dt^2+\fr{dr^2}{\Big(1-{\l^2}{r^2}\Big)}
     +r^2(d\th^2+\sin^2\th d\f^2)
\eea
As well known, this de Sitter spacetime has a cosmological horizon at $r=\fr1{\l}$ which is related to the cosmological constant by
\bea
r_{c.h.}\equiv \fr1{\l}\equiv \sqrt{\fr{3}{\L}}
\eea
The physical states are annihilated by the lapse function constraint by definition: 
\bea
\Big[\cR+K_{mn}K^{mn}-K^2   -2\L\Big]|phys>=0  \la{nconstrq}
\eea 
This condition implies that the bulk degrees of freedom are in vacuum. 
In \cite{Higuchi:1991tk} a similar observation was made at a linear level. 
The boundary degrees of freedom may be excited but they must be invariant under translation along the $r$-direction. 
Let us focus on the ``boundary" degrees of freedom. To understand how to construct such states let us turn to the following global coordinates related to the static ones by
\bea
\tilde{t} &=& t+\fr1{2\l}\ln (1-{\l^2}r^2)\nn\\
\tilde{r} &=& \fr{r}{e^{\l t+\fr1{2}\ln (1-{\l^2} r^2)}} \la{newco1}
\eea
which turns the metric into
\bea
ds^2 &=& -d\tilde{t}^2+e^{2\l\tilde{t}}\Big[d\tilde{r}^2+\tilde{r}^2(d\th^2+\sin^2\th d\f^2) \Big]  \la{mnewco02}
\eea
The inverse relations of the two coordinate systems are given by
\bea
r &=& \tilde{r} e^{\l\tilde{t}} \nn\\
t&=&\tilde{t}-\fr1{2\l}\ln\Big(1-\l^2\tilde{r}^2e^{2\l\tilde{t}}\Big)   \la{newco2}
\eea
As $r\ra \fr1{\l}$ the new coordinates $\tilde{t}, \tilde{r}$ take asymptotic values. All of these are standard. The idea now is to construct the boundary states in the asymptotic region of $\tilde{t}, \tilde{r}$ coordinates and transcribe them into the original coordinates through a Bogoliubov transformation. (We will come back to this in the conclusion.) The locations of the operators in the asymptotic region in the $\tilde{t}, \tilde{r}$ coordinates will be invariant under a finite translation and the states constructed out of such operators are thus invariant. We should consider the translation in $\tilde{r},\tilde{t}$ that shifts only $r$ but not $t$ (or the angular coordinates); this translation corresponds to a shift in $r$ in its ``asymptotic region," $r\simeq \fr1{\l}$.

Let us review how the infrared divergence arises in the two-point function calculated in
the conventional setup.
For the perturbative analysis one should consider the metric field equation of \rf{kin} for the mode expansion \cite{Higuchi:1986ww}\cite{Higuchi:1991tk}\cite{Bernar:2014lna}. Since the metric tensor structure is irrelevant for the present discussion, let us suppress the indices and denote the metric by $\Phi$ instead: 
\bea
\nabla_0^\m \nabla_{0\m} \Phi=0
\eea
where $\nabla_0^\m$ denotes the covariant derivative with the dS background metric. 
Letting the solution of the scalar field equation equal
\bea
\Phi_{\w lm}(t,r,\th,\vf)=c^{\w l}\,e^{-i\w t}Y_{lm}(\th,\vf)u_\w(r)
\eea
where $c^{\w l}$ is a normalization factor and $Y_{lm}$ denotes the usual spherical harmonics. The quantum number $\w$ enumerates the ``radial wave function," $u_\w$, counting the degrees of freedom associated with the $r$-direction. 
The radial wave function $u_\w$ satisfies
\bea
\Big[\fr{1}{r^2}\pa_r(r^2(1-\l^2 r^2)\pa_r)+\fr{\w^2}{1-\l^2 r^2}-\fr{l(l+1)}{r^2}\Big]u_\w=0
\eea
The solution regular at $r=0$ is given by
\bea
u_\w=r^l\Big(1-\fr{\L}{3}r^2\Big)^{\fr{i\w}{2}\sqrt{\fr{3}{\L}}}
{}_2F_1\Big(\fr{l}{2}+\fr{i\w}{2}\sqrt{\fr{3}{\L}},\fr32+\fr{l}2+\fr{i\w}{2}\sqrt{\fr{3}{\L}},\fr32+l,\fr{\L}{3} r^2\Big)
\eea
Consider the mode expansion
\bea
\Phi=\int d\w \sum_{lm}\Big[c^{\w l} u_\w(r)Y_{lm}e^{-iwt}\,a_{\w lm}+h.c.\Big]
\eea
where $c^{\w l}$ is a normalization factor and the annihilation operator has been labeled by $a_{lm}$. The two-point amplitude is given by
\bea
<0|\Phi(x)\Phi(x')|0>=\sum_{l,m}\int_0^\infty d\w \Big[
\fr{\bar{\Phi}_{\w lm}(x)\Phi_{\w lm}(x')}{e^{2\pi \w}-1}
-\fr{{\Phi}_{\w lm}(x)\bar{\Phi}_{\w lm}(x')}{e^{-2\pi \w}-1}
\Big]
\eea
The infrared divergence comes from the $l=0$ mode in the $\w\ra 0$ region (see \cite{Bernar:2014lna} for more details). 

The infrared divergence does not arise in the present setup due to the fact that the physical state condition eventually leads to $\w=0$ as we now show. The physical state condition requires that the physical states be invariant under translation along the $r$-direction. This can be assured by considering operators at an ``asymptotic" region. In the Schwarzschild case the ``boundary" operators $\Phi(r)$ (the rest of the coordinates have been suppressed) have asymptotically large values of $r$ \footnote{Although the consideration of a large value of $r$ was motivated from a different direction in \cite{Park:2015xoa}, the motivations there and in the present work are consistent, and everything seems to hold together.} and thus the location of the operator remains invariant under finite translation along $r$. For the dS case, it is more subtle: the boundary is $r=r_{c.h.}$, the cosmological horizon. There are several indications that this choice/interpretation is a valid one. Firstly, as we have shown, the $r\sim \fr1{\l}$ region corresponds to an asymptotic region in the global coordinates wherein the boundary states are readily constructed.
Moreover, although the coordinate distance of $r$ is not infinite it may be safe to view the location of the horizon as practically infinite. The physical distance to the horizon is large due to a small value of the cosmological constant and therefore it should be possible to practically disregard a relatively small finite change in the physical distance of the operator. Secondly, it was proposed \cite{Banks:2000fe} in the context of dS/CFT that the cosmological horizon may be taken as the location of the holographic screen. Although our choice of $r=r_{c.h.}$ has been motivated by an unrelated reason we also believe that the resulting boundary ``theory" must be suitable for describing the physics of the static observer.

With the boundary set at the horizon, let us examine the equation for $u_\w$ which we quote for convenience:
\bea
\Big[\fr{1}{r^2}\pa_r(r^2(1-\l^2 r^2)\pa_r)+\fr{\w^2}{1-\l^2 r^2}-\fr{l(l+1)}{r^2}\Big]u_\w=0;
\eea
by multiplying ${1-\l^2 r^2}$ it can be rewritten as
\bea
\Big[({1-\l^2 r^2})\fr{1}{r^2}\pa_r(r^2(1-\l^2 r^2)\pa_r)+{\w^2}-({1-\l^2 r^2})\fr{l(l+1)}{r^2}\Big]u_\w=0
\eea
Once the $r\ra \fr1{\l}$-limit is taken, only the $\w$-containing term survives, thus leading to\footnote{Or $u_0=0$ but this case can be included in $\w=0$ case.}
\bea
\w=0\;\;
\eea 
Therefore considering the ``boundary" region $r\ra \fr1{\l}$ indeed implies $\w=0$.
The fact that the infrared divergence does not arise is now obvious.\footnote{ 
Curiously, with $\w=0$ $a_{\w=0, l=0, m=0}$ commutes with the rest of the operators and therefore completely drops out of the amplitude computations.}  
For the perturbative analysis one can follow the manifestly 3D approach of \cite{Park:2014tia} and \cite{Park:2014noa}.

With regards to the work of \cite{Allen:1985ux} mentioned in the beginning of this section, an observation was made therein on the reduction of the full de Sitter symmetry to $E_3$, the isometry of the 3D Euclidean space. 
Its meaning can be appreciated in the context of the present work: $E_3$ would be the isometry symmetry resulting from the choice of  $x^3$ to be $\tilde{t}$ in the global coordinate system \rf{mnewco02} and carrying out the present reduction scheme.

%%%%%%%%%%%%%%%%%%%%%%%%%%%%%%%%%%%%%
%%%%%%%%%%%%%%%%%%%%%%%%%%%%%%%%%%%%%
\section{Conclusion}
%%%%%%%%%%%%%%%%%%%%%%%%%%%%%%%%%%%%%
%%%%%%%%%%%%%%%%%%%%%%%%%%%%%%%%%%%%%

In this work we have extended the quantization scheme proposed in \cite{Park:2014tia,Park:2014qoa} and further developed in \cite{Park:2015qxa,Park:2014noa,Park:2015ota,Park:2015xoa} to gravity matter systems and pure dS gravity. 
The shift vector constraint can either be explicitly solved with gauge-fixing of the shift vector. The Hamiltonian becomes identical to the lapse function constraint that serves as the physical state condition, and this was one of the key elements leading to the reduction of the physical states.
The connection between the $U(1)$ algebra of the totally geodesic foliation and the $x^3$-evolution operator has been made more specific as well. A more elaborate gauge-fixing procedure was necessary to establish renormalizability in the presence of the matter fields. 
As for the dS gravity, the intermediate 4D covariant method could not, unlike the flat or Schwarzschild case, be employed because of the occurrence of the infrared divergence in the 4D setup.  
A novel interpretation of the lapse function constraint in conjunction with the boundary states was an additional key element for the reduction (see footnote \ref{fninfty} in this regard). 
The well-known infrared divergence does not arise in the present scheme of quantization because the reduction in effect removes the mode sum over the radial modes that induced the divergence in the conventional setup.

In addition to the complexities of a curved background, the dS case has revealed the subtle issue of the boundary degrees of freedom. We have proposed that invariance under the $x^3$-translation 
narrows the physical sector of the theory to the states that can be constructed out of the operators located at the cosmological event horizon. The construction of the physical states is facilitated by going to the global coordinate system \rf{newco2}.
There have been intensive debates over existence of a dS invariant vacuum   \cite{Bernar:2014lna,Tsamis:1996qq,Janssen:2008px,Marolf:2010nz} (and the refs therein). It appears that the reduction of the isometry observed in \cite{Allen:1985ux} can be given a clearer meaning in the present context: the dS symmetry is reduced due to the reduction of the physical states.

\vspace{.2in}

There are several future directions:

One may wonder why the present quantization scheme is only applicable to certain geometries with relatively simple foliation. In general, choosing the right degrees of freedom in a gravity theory is important. (The relevance and importance of foliation (or slicing) was discussed, e.g., in \cite{Ghezelbash:2002vz} in the context of (A)dS/CFT correspondence.)
It will be worthwhile to explore in a mathematically rigorous manner the most general forms of a metric to which the present quantization scheme can be applied. 
Once quantization is successfully carried out in one coordinate system, it should be possible to translate the results into another coordinate system; the conversions between the static coordinate result and a global coordinate result 
should be implemented by Bogoliubov transformations and it will be of some interest to work out complete details. (A work on the Bogoliubov transformation can be found in \cite{Sato:1994ac}.)

\vspace{.2in}
By the gauge-fixing of a matter component the original scalar can appear as one of the graviton mode. Although this gauge-fixing has brought along the renormalizability, the usefulness of such a description in other contexts remains to be explored. 
This is particularly true because the description leads to a rather radical-sounding question of whether a scalar may be viewed as a graviton depending on how one describes it. Such a description may possibly be viewed analogous to an increase of the polarization degrees of freedom of a vector field after it has eaten up a scalar in the standard gauge theory.
To explore the issue in a setup applicable to, say, inflation physics, one may consider a system of metric coupled with a scalar but now with a potential. It would be of some interest to work out the details of renormalization procedure in the setup in which the scalar is treated as a fixed background after the fluctuation is gauged away and to study its implications for the inflationary context.

\vspace{.2in}
Another direction, not unrelated to the previous one, is to investigate whether or not renormalizability could be established without the mechanism of converting the matter component(s) into a metric component.
In the conventional context, it is the graviton degrees of freedom that get in the way of renormalizability. In this sense it is ironic that in the present method the components of matter fields are traded for the additional graviton degrees of freedom in order for the renormalizability. On one hand this seems to reflect the importance of choosing the right degrees of freedom in a gravity theory. On the other hand the usefulness of the mechanism is rather limited. For example, if the gravity-scalar system has more than two scalars it will be inevitable to directly deal with the matter fields at the dynamical level. 

However,
we hope that the present approach will be useful for the following purpose: the renormalizability established in this work may hint towards a possibility for renormalizability even in terms of the original (i.e., without conversion into metric degrees of freedom) matter degrees of freedom through potentially complicated field redefinition(s). 
This definitely is a possibility for the matter-gravity systems considered in this work. (Recently progress has been made in this direction; see \cite{Park:2016fxc}.)
Even if renormalizability can be carried out in terms of the original fields, highly nontrivial field redefinitions are likely to be required. Finding the necessary field redefinitions without knowing their origin would be prohibitively difficult. However, if one knows a set of the degrees of freedom with which renormalizability is established, it may be possible to deduce the forms of the field redefinitions with relatively less efforts.   

One of the most challenging and important tasks will be to (dis)prove the renormalizability of a more general gravity-matter system such as a gravity-scalar-vector system. It will presumably be possible to establish reduction along the lines of the present analysis. However, to see whether that will be sufficient for renormalizability would require one to examine subtle conceptual issues and carry out highly (and unpleasantly) technical analyses. 
 The effective action resulting from integrating out some of the high energy modes would have different characteristics depending on whether one has the reduction (or not, as in the conventional setup). With the reduction it will be possible to replace all the Riemann curvature tensors by a combination of the 3D Ricci tensor and metric. In turn, that will allow one to replace all the Ricci tensor/scalar terms by the matter fields through the metric field equation. The implications of this feature remain to be explored. 
 
Finally, we comment on the following very speculative yet intriguing possibility. 
In contrast to the reduction considered in \cite{Park:2013vpa,Park:2013bma} where {\em manual} reduction was considered, the reduction of \cite{Park:2014tia,Park:2014noa,Park:2015xoa,Park:2015ota} and the present work is ``spontaneous" in the sense that it takes place as a result of gauge-fixing rather than narrowing down to a certain lower dimensional sector of the starting theory. In spite of this difference, the reduction scheme of \cite{Park:2014tia} is expected to share a novel feature of the manual reduction of \cite{Park:2013vpa}, the relevance of the so-called ``virtual boundary" terms. These boundary terms were introduced in \cite{Park:2013vpa,Park:2013bma} in order to ensure the consistency of the reduction and should be viewed as part of the specification of the theory. In a similar manner, it seems to be a reasonable possibility - radical though it may sound - that one may add the virtual boundary terms in the starting bulk action in such a way as to cancel the loop divergences of the reduced 3D theory at a later stage. 
If the presence of such boundary terms could be fully justified, which would require more work, that would take us very close to establishing renormalizability of a general gravity-matter system.

We will report on progress in some of these issues in the near future.

\vspace{.7in}
\ni {\bf Acknowledgment}\\

\ni I thank Hyun Seok Yang at KIAS, Seoul South Korea for sharing his expertise on various aspects of quantum gravity.

\newpage
\appendix

\renewcommand{\theequation}{A.\arabic{equation}}
 \setcounter{equation}{0}
\section{Conventions and identities}

The signature is mostly plus: 
%%%
\bea
\eta_{\m\n}=(-,+,+,+)
\eea
%%%
All the Greek indices are four-dimensional
%%%
\bea
\a,\b,\g,...,\m,\n,\r...=0,1,2,3
\eea
%%%
and all the Latin indices are three-dimensional
%%%
\bea
a,b,c,...,m,n,r...=0,1,2
\eea
%%%
The complete set of the bulk de Donder gauge $g^{\r\s}\G^\m_{\r\s}=0$ \cite{Smarr:1978dia} is given by
%%%
\bea
&& (\pa_{x^3}-N^m \pa_m) n=n^2K  \nn\\
&& (\pa_{x^3}-N^n \pa_n)N^m=n^2(\g^{mn}\pa_n \ln n-\g^{pq}\G^m_{pq})
\eea
%%%
in the ADM fields. The 3D Ricci scalar, Ricci tensor and Riemann tensor are denoted respectively by
\bea
\cR,\cR_{mn},\cR_{mnpq}
\eea
The fluctuation metric $\f_{\m\n}$ is defined through
\bea
g_{\m\n}\equiv g_{0\m\n}+\f_{\m\n}
\eea
%%%
The indices of $\f_{\m\n}$ are raised and lowered by $g^{0\m\n},g_{0\m\n}$.
The following shorthand notations were used in some places:
\bea
\f\equiv g^{0\m\n}\f_{\m\n}\quad,\quad \f^\m\equiv \pa_\k\f^{\k\m}
\eea
%%%
The graviton propagator is given by
%%% 
\bea 
<\f_{\m\n}(x_1)\f_{\r\s}(x_2)>&=& P_{\m\n\r\s}\, \D(x_1-x_2) 
\eea
%%%
where, for the traceless propagator\footnote{{Imposition of the traceless condition with the de Donder gauge was previously mentioned in \cite{Ortin} as we have recently become aware of.}},
%%%
\bea
P_{\m\n\r\s}\equiv \fr12\Big(g_{0\m\r}g_{0\n\s}+g_{0\m\s}g_{0\n\r}
              - \fr12g_{0\m\n}g_{0\r\s}\Big);
\eea
%%%
The 3D fluctuation metric is introduce through
\bea
h_{mn}\equiv h_{0mn}+q_{mn}
\eea
The second fundamental form splits accordingly:
\bea
K_{mn}=K_{0 mn}+k_{ mn}
\eea
where $K_{0 mn}$ denotes the classical value and 
\bea
k_{mn}\equiv
\fr1{2n}\mathscr{L}_{\pa_{x^3}} q_{mn}
\la{K4defqqlinq}
\eea
%%%
after $N_m=0$ gauge.

\renewcommand{\theequation}{B.\arabic{equation}}
 \setcounter{equation}{0}
\section{Gauge-fixing of scalar field}

In this appendix we elaborate on the procedure of gauge-fixing of the scalar field whose result has been used in section 3.1.
As noted in the main body the system reduces to 3D once the shift vector constraint is enforced and the residual diffeomorphism with 3D coordinate dependence can be employed. We employ the setup of the background field method.

Consider a generic scalar field $\s(x)$:
\bea
\s(x):\quad \mbox{a generic scalar field}
\eea
Its infinitesimal and finite transformations respectively are 
\bea
\s'(x)=\s(x)+\xi^\m \pa_\m \s(x)  \la{gr}
\eea
and (see, e.g., \cite{GPP})
\bea
\s'(x)=e^{\mathscr{L}_\xi}\s(x)
\eea
where $\mathscr{L}_\xi$ denotes the Lie derivative along the vector field $\xi$. 
We start with the gauge-fixing of the scalar in a general background.\footnote{
The following naive procedure leads to a problem in the flat case in which the scalar solution is set to 
\bea
\z_0=0
\eea 
Consider shifting $\z=\z_0+\tilde{\z}$. Considering $\z'=\z_0+\tilde{\z}'$ and gauging away $\tilde{\z}'$ one gets, at the infinitesimal level,
\bea
\z'(x)=\z(x)+\xi^\m \pa_\m \z(x)
\eea
The problem becomes apparent once we consider the finite gauge transformation:
\bea
\z'=e^{\mathscr{L}_\xi}\z=\z_0
\eea
where $\mathscr{L}_\xi$ denotes the Lie derivative along the vector field $\xi$. This implies, since $\z_0=0$,
\bea
\z=e^{-\mathscr{L}_\xi}\z_0=0
\eea
which is a contradiction. }
In the background field method (a review of the background field method tailored for the present context can be found in \cite{Park:2015ota}) one shifts the original field to
\bea
\s\equiv \s_0+\hat{\s}
\eea
where $\s_0$ denotes the classical solution and $\hat{\s}$ the fluctuation. For the reason explained in \cite{Park:2015ota} let us introduce another shift $\hat{\s}= \bar{\s}+{\tilde{\s}}$ and consider
\bea
\s\equiv \s_B+{\tilde{\s}} \quad,\quad  \s_B \equiv  \s_0+{\bar{\s}}
\eea
where $\tilde{\s}$ is taken as the fluctuation field.
When computing Feynman diagrams the $\s_B$ fields are placed one the external lines.
Coming back to the present case, let us define
\bea
\z\equiv \z_B+\tilde{\z}
\eea
The field $\z$ transforms according to the general rule \rf{gr}:
\bea
\z'=\z_B+\tilde{\z}+\xi^\m \pa_\m (\z_B+\tilde{\z})
\eea
``Transferring" the entire transformation to the fluctuation part (see e.g. \cite{'tHooft:1974bx,Deser:1974cz}), one gets
\bea
\z'\equiv \z_B+\tilde{\z}'
\eea
Matching the two equation leads to
\bea
\z_B+\tilde{\z}'=\z_B+\tilde{\z}+\xi^\m \pa_\m (\z_B+\tilde{\z})
\eea
Let us gauge away  $\tilde{\z}'$; setting $\tilde{\z}'=0$ implies
\bea
\xi^\m \pa_\m \z_B+\tilde{\z}+\xi^\m \pa_\m \tilde{\z}=0
\eea
Therefore it should be possible to gauge away $\tilde{\z}'$ by solving this equation. 
With $\tilde{\z}$ gauged away, the Feynman rules should be adopted accordingly. First of all, the Feynman diagrams with internal scalar loops become irrelevant. Secondly, although the $\z_B$ fields are formally placed on the external lines as before one can now set $\bar{\z}=0$ and thus identify $\z_B=\z_0$.  (More carefully one should consider
\bea
\z\equiv \z_B(x)+\tilde{\z}(y)
\eea 
and gauge away $\tilde{\z}(y)$. The argument $y$ of $\tilde{\z}(y)$ has been explicitly recorded to emphasize the 3D nature of the fluctuation $\tilde{\z}$. 
)

The discussion so far was for an arbitrary coordinate-dependent background. For a flat case we use the ``analytic continuation," namely, set $\z_0=0$, at the end. In effect the whole procedure then amounts to the assertion \rf{asser}.

\newpage
%%%%%%%%%%%%%%%%%%%%%%%%%%%%%%%%%%%%%%%%%%%%%%%%%%%%%%%%%%%%%%%%

\end{document}